\documentclass[journal=jctcce,manuscript=article,layout=traditional]{achemso}
\usepackage{helvet}
\usepackage{booktabs}
\usepackage{xcolor}
\usepackage{amsmath}
\usepackage{enumitem}

\usepackage[utf8]{inputenc}
\usepackage{url}

\newcommand{\etal}{\emph{et\,al.}}
\newcommand{\Mg}{Mg$^\mathrm{2+}$}

\title{Evolution of protein--RNA interactions}

\author{Michal H. Kolář}
\affiliation{Department of Physical Chemistry, University of Chemistry and Technology, Technická 5, 16628 Prague, Czech Republic}

\author{Klára Hlouchová}
\affiliation{Department of Cell Biology, Faculty of Science, Charles University, Viničná 7, 12843 Prague, Czech Republic}
\alsoaffiliation{Institute of Organic Chemistry and Biochemistry, Czech Academy of Sciences, Flemingovo nám. 2, 16610 Prague, Czech Republic}
\email{klara.hlouchova@natur.cuni.cz}

\begin{document}

\maketitle

\begin{abstract}
\singlespacing
Since the Hadean era of Earth's history, peptides/proteins and RNA have undergone a complex evolutionary trajectory. Originating from simple monomeric units, these molecules evolved abiotically under various biochemical and biophysical constraints into functional biomolecules that contributed to the emergence of the first living cells. Within these cells, their interactions could then evolve through Darwinian selection. In this review, we examine current understanding of how protein--RNA interactions emerged under prebiotic conditions and developed into today’s iconic biomolecular machines such as the ribosome. Particular emphasis is placed on the types of physicochemical interactions accessible to early protein--RNA complexes and their roles in driving spatial organization and compartmentalization in protocellular environments.
\end{abstract}

\singlespacing

\section*{Introduction}

Protein--RNA interactions play an essential role across diverse cellular functions, from modulating transcript stability to operating the assembly of biomolecular machines such as ribosomes and spliceosomes. This functional breadth is mirrored by the structural diversity of RNA itself, which---unlike DNA---offers a rich array of recognition elements through its flexible backbone, varied base geometries, and the presence of 2'-hydroxyl groups (\texttt{1}). These structural features enable proteins to interact with RNA through multiple, often cooperative, modes of binding, supporting both transient contacts and highly stable assemblies.

\begin{figure}[tb]
    \centering
    \includegraphics{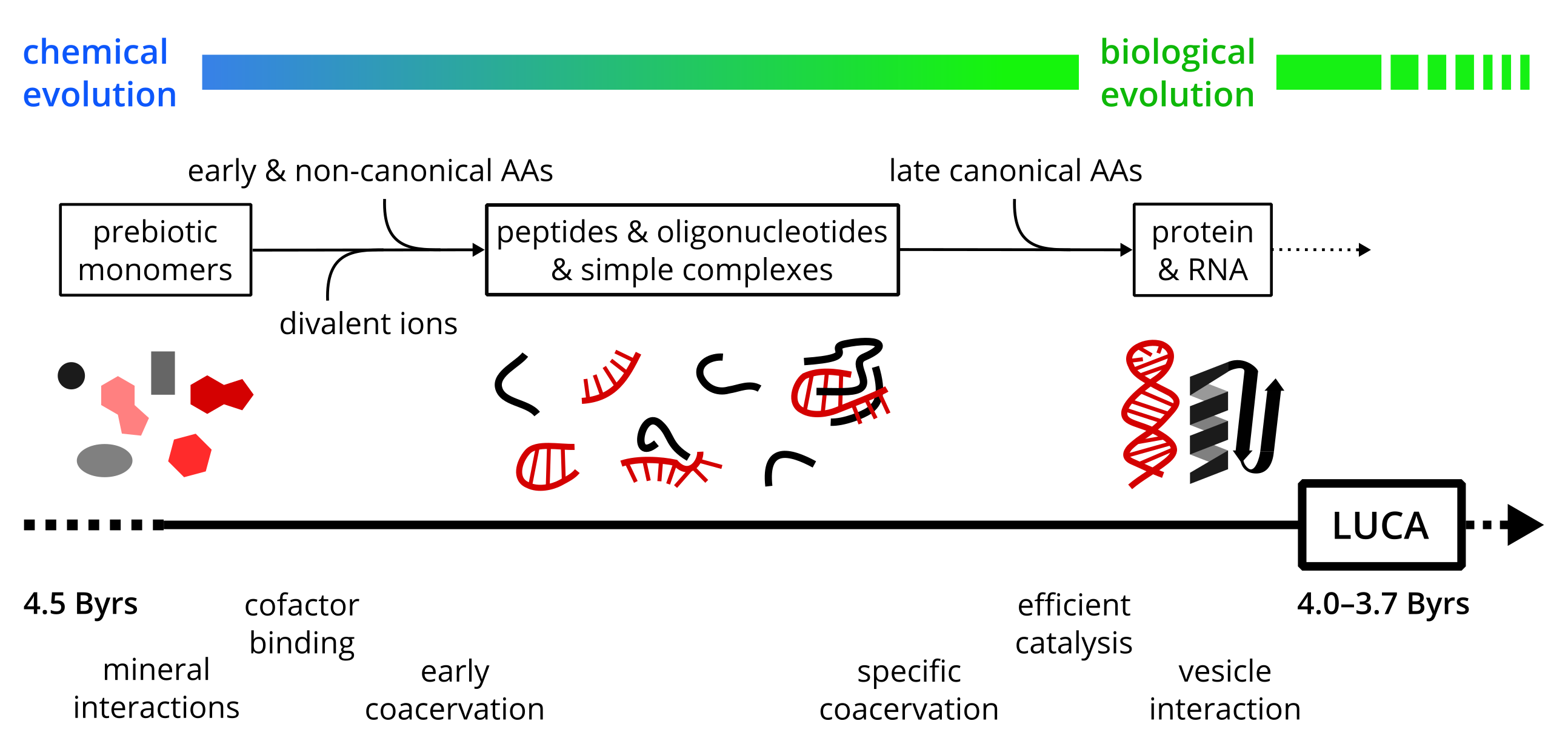}
    \caption{Peptide/protein-RNA interactions probably date back to Hadean era of Earth’s history about 4.5 billion years (Byrs) ago. Recent experimental data provide support to hypotheses that these interactions could take place before the evolution of the full protein alphabet, in the absence of canonical basic and aromatic residues (\texttt{6,8,43}). At the same time, canonical basic residues would substantially improve protein--RNA assembly, coacervation, and catalysis (\texttt{11,53}).}
    \label{fig:timeline}
\end{figure}

Evidence from prebiotic chemistry suggests that the molecular components of both peptides and RNA were present among early Earth organics (\texttt{2-4}). While opinions differ on the relative ease of prebiotic polymerization, stability, and abundance of peptides versus polynucleotides, there is broad consensus that both polymer types coexisted in the earliest stages of life’s evolution (\texttt{5-11}). Alongside mineral, lipids, metals and small-molecule cofactors, these polymers likely contributed to a chemical network that led to the emergence of proto-biological organization (\texttt{5,9,10}). Prior to the emergence of templated synthesis or enzymatic replication, early interactions between short, compositionally biased peptides and RNAs may have laid the groundwork for the functional integration that now defines the central dogma. Understanding how these interactions evolved from chemically driven associations into the specific, versatile roles observed today remains a key question at the intersection of structural biology and the origins of life (Fig.\,\ref{fig:timeline})

\section*{Ancient RNA-binding themes}

The “RNA world” hypothesis traditionally casts peptides as cofactors in an RNA-centric origin of life (\texttt{12,13}). While certain peptide motifs were likely selected for their RNA-binding properties (as described below), this review takes a different stance, placing peptides rather as equal contributors in a systems-chemistry view of life’s origins (Fig.\,\ref{fig:timeline}). 

The current protein structure space has been largely shaped by ancestral constraints, with fewer than ten distinct folds responsible for over 30\% of all entries in the Protein Data Bank (PDB) (\texttt{14}). This has led to the proposition that many of today’s folds arose by recombinations and fusions of ancient peptide ``vocabulary'' of supersecondary structures (\texttt{11,13,15}). In this context, Kocher and Dill proposed that peptides displaying (i) intrinsic folding stability and resistance to degradation, and (ii) the ability to participate in autocatalytic networks, would be favored and propagated through evolution-like processes (\texttt{16}). Supporting this hypothesis, recent analyses have identified relatively short polypeptide fragments---approximately 10 to 80 residues in length---that are shared among structurally unrelated domains (\texttt{17,18}). About half of the fragments correspond to common supersecondary structure elements, about 20\% are associated with metal binding (\texttt{17}), collectively suggesting selection based on folding and catalytic functionality (\texttt{19}). Notably for the focus of this review, about 40\% of these shared motifs do not form compact folded structures and are instead associated with nucleotides, RNA, or DNA binding. About 25\% are found in ribosomal proteins. These findings support the idea that such motifs may represent molecular relics of primordial peptide–RNA co-evolution.

\section*{History or protein-RNA interactions inferred from the ribosome}

Much of our current understanding of the evolution of protein–RNA interactions comes from studies of the ribosome (\texttt{20}), the sophisticated protein--RNA complex responsible for protein synthesis in all known life. The core structure of the modern ribosome predates the Last Universal Common Ancestor (LUCA), making it a molecular archive of early selection processes. As such, the ribosome captures key milestones in the co-evolution of proteins and RNA, not only as shaped by Darwinian selection across the three domains of contemporary life (\texttt{21}), but likely also by prebiotic biophysical and biochemical constraints that preceded biological evolution. 

A detailed model of the ribosome’s evolution has been developed, providing a chronological framework for the accretion of its structural layers (\texttt{22-24}). The independent abiotic evolution of the small ribosomal subunit (SSU) and large ribosomal subunit (LSU) eventually transitioned into coordinated co-evolution, culminating in the emergence of the full translation apparatus, including mRNA, tRNAs, and translation factors, by the time of LUCA. According to the model, a minimal protoribosomal RNA (prRNA) of the LSU gradually increased in length and complexity.

\begin{figure}[tb]
    \centering
    \includegraphics[width=\textwidth]{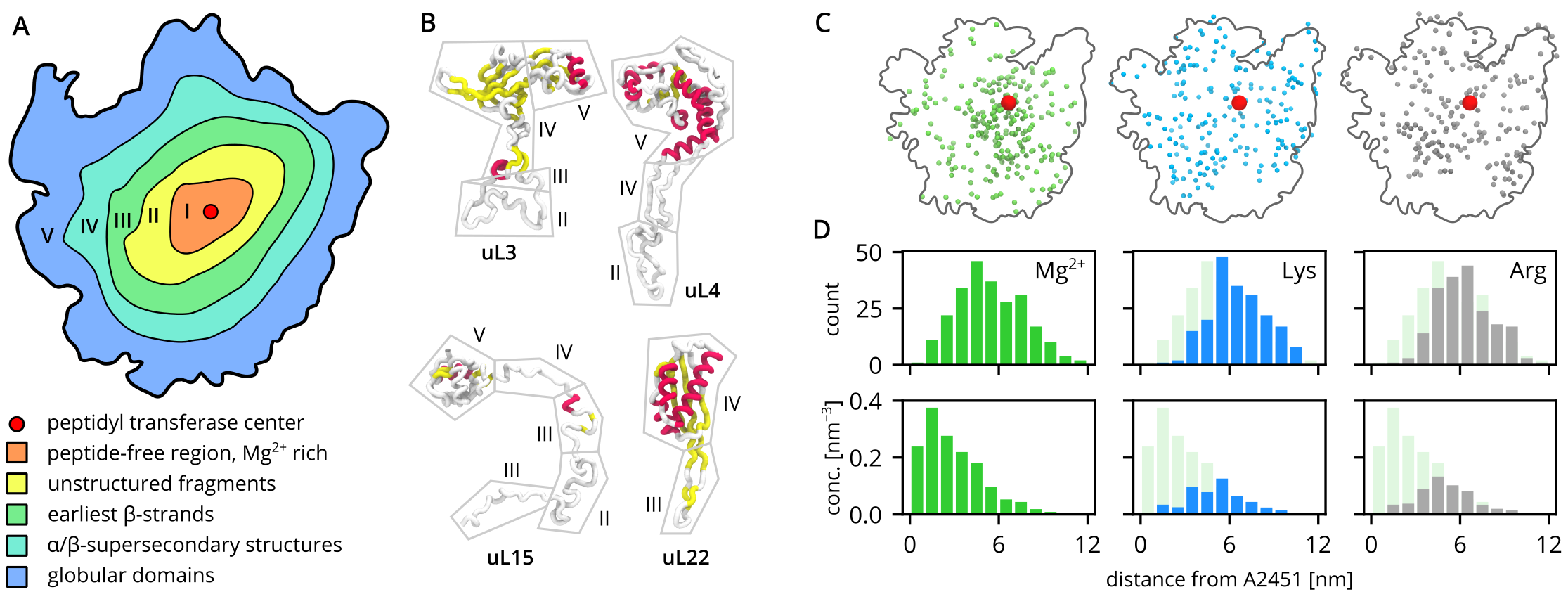}
    \caption{A) Schematic representation of r-protein evolution in the LSU, as described in Kovacs \etal (\texttt{27}). B) Four representative r-proteins with $\alpha$-helices in red and $\beta$-strands in yellow. Evolutionary fragments following the model proposed by Kovacs \etal (\texttt{27}) are shown. Regions II--V correspond to Phases 3 to 6 in the same reference. C) Spatial distribution of \Mg \,ions (green) and the side chains of Lys (blue) and Arg (gray), shown relative to LSU rRNA. The nucleotide A2451 at the peptidyl transferase center (PTC) is highlighted as a red sphere. D) A distribution of \Mg \,ions and Lys and Arg side chains as a function of distance from A2451 at the PTC. The upper panels show absolute counts within spherical shells; the lower panels show the corresponding number concentrations. \emph{E.\,coli} ribosomal structure PDB 5AFI was used for the analysis.}
    \label{fig:ribosome}
\end{figure}

An unprecedented look into the history of protein–RNA interactions is made possible by the characteristic architecture of r-proteins and their phylogenetic relationships within modern ribosomes. The bacterial ribosome contains approximately 57 r-proteins, 34 of which have homologs in higher organisms and are therefore considered universal (\texttt{25}). These proteins typically exhibit a multi-domain structure, with globular domains positioned near the ribosome surface and elongated extensions reaching deep into the ribosomal core. Both terminal extensions and hairpin loop structures are commonly observed (Fig.\,\ref{fig:ribosome}B). 

Several studies have classified fragments of r-proteins into chronological categories based on the presumed timing of their incorporation into the protoribosome (\texttt{26,27}) with the most ancient parts buried deep in the ribosome (Fig.\,\ref{fig:ribosome}A). The oldest fragments are found in seven universal r-proteins of LSU, including uL2 and uL3. A notable feature of these ancient fragments is their lack of defined secondary structural motifs, suggesting a simpler, more flexible, and less specific interaction with prRNA. Our recent work supports this hypothesis. We investigated the interaction of two prRNA constructs---small (136nt) and large (617nt)---with fragments of r-proteins (\texttt{28}). Using microscale thermophoresis and atomistic computer simulations, we found that the peptides bind more specifically to the larger prRNA construct, indicating increased structural specificity with RNA
complexity.

Moving outward from the center of the LSU toward its surface, the propensity for secondary structure elements increases, and supersecondary structures become more common (\texttt{26}). At the ribosomal periphery, folded domains begin to appear, although these initially lack hydrophobic cores. This suggests that early protein--RNA interactions involved largely unstructured peptides, with more structured peptides and proteins emerging only after undergoing biophysical optimization (\texttt{29}).

A recent study proposes that, in later stages of ribosome evolution, the optimization of protein--RNA interactions was driven by the need to allosterically coordinate distant regions of the ribosome (\texttt{30}). This phase coincided with the recruitment of aromatic residues into r-proteins at the protein--RNA interface. The resulting allosteric information transfer facilitates ribosomal dynamics during the tRNA translocation (\texttt{31}) and the association with translation factors (\texttt{32}).

Like other RNA-binding proteins, r-proteins are particularly rich in the basic residues Arg and Lys. However, the oldest regions of the ribosome are devoid of proteins, and the negative charge of the rRNA backbone in these areas is instead neutralized by \Mg \,ions (Fig.\,\ref{fig:ribosome}CD) (\texttt{33}). This observation suggests that early peptides may have gradually replaced ion--RNA interactions, thereby enhancing the specificity of RNA charge compensation.

\section*{Early ionic interactions}

Extant protein--RNA interactions involve a wide range of canonical amino acids, most frequently the positively charged Arg and Lys, as well as aromatic, other non-polar (Gly, Ile, Met) and polar (Ser, Asn) residues. These interactions encompass the full spectrum of possible binding modes -- including hydrogen bonds, van der Waals and stacking interactions, and general hydrophobic effects (\texttt{1}). Based on our current understanding of prebiotic chemistry, it is widely accepted that not all canonical amino acids would have been available to prebiological peptides, early proteins, and primordial metabolism. A consensus subset---often referred to as the ``early'' alphabet---comprises approximately half of the canonical set (Gly, Ala, Asp, Val, Glu, Ile, Leu, Pro, Ser, and Thr) (\texttt{34}. These amino acids, along with other non-canonical variants, have been consistently detected in multiple sources of spontaneous abiotic synthesis (\texttt{34-36}). These include extraterrestrial sources (e.g. meteorites), experiments using simple gases under both reducing and non-reducing conditions, and natural energy inputs such as lightning and volcanic activity. Although feasible prebiotic routes for synthesizing other canonical amino acids---such as Arg and Cys---have been proposed, their availability was likely limited (\texttt{4,5}). This is due to the narrow set of conditions compatible with their synthesis and the requirement of multiple sequential reactions. This stands in contrast with the ubiquitous presence of the early amino acid set.

Several researchers have questioned the plausibility of protein--RNA interactions in a prebiotic context, given the likely scarcity of basic and aromatic amino acids in early peptides/proteins (\texttt{37,38}). To directly address the biophysical constraints of protein--RNA interactions relevant in prebiotic scenarios, Blanco \etal \,analyzed protein--aptamer complexes that have not been shaped by eons of other evolutionary pressures (\texttt{37}). They found that basic and aromatic residues were dominant in these interactions, while small hydrophobic amino acids were under-represented. Blanco \etal \,concluded that the absence of Arg or a biophysically similar amino acid would render early protein--RNA interactions improbable (\texttt{37}). However, it should be noted that the protein--aptamer complexes were results of \emph{in vitro} evolution studies starting from biological protein templates and the bias towards basic and aromatic residues could partly be introduced upon creation of the starting libraries. Contrasting with the conclusions drawn by Blanco \etal, two alternative models have been proposed for protein--RNA interactions in the absence of canonical basic amino acids (\texttt{39}).

\begin{figure}[t]
    \centering
    \includegraphics[width=\textwidth]{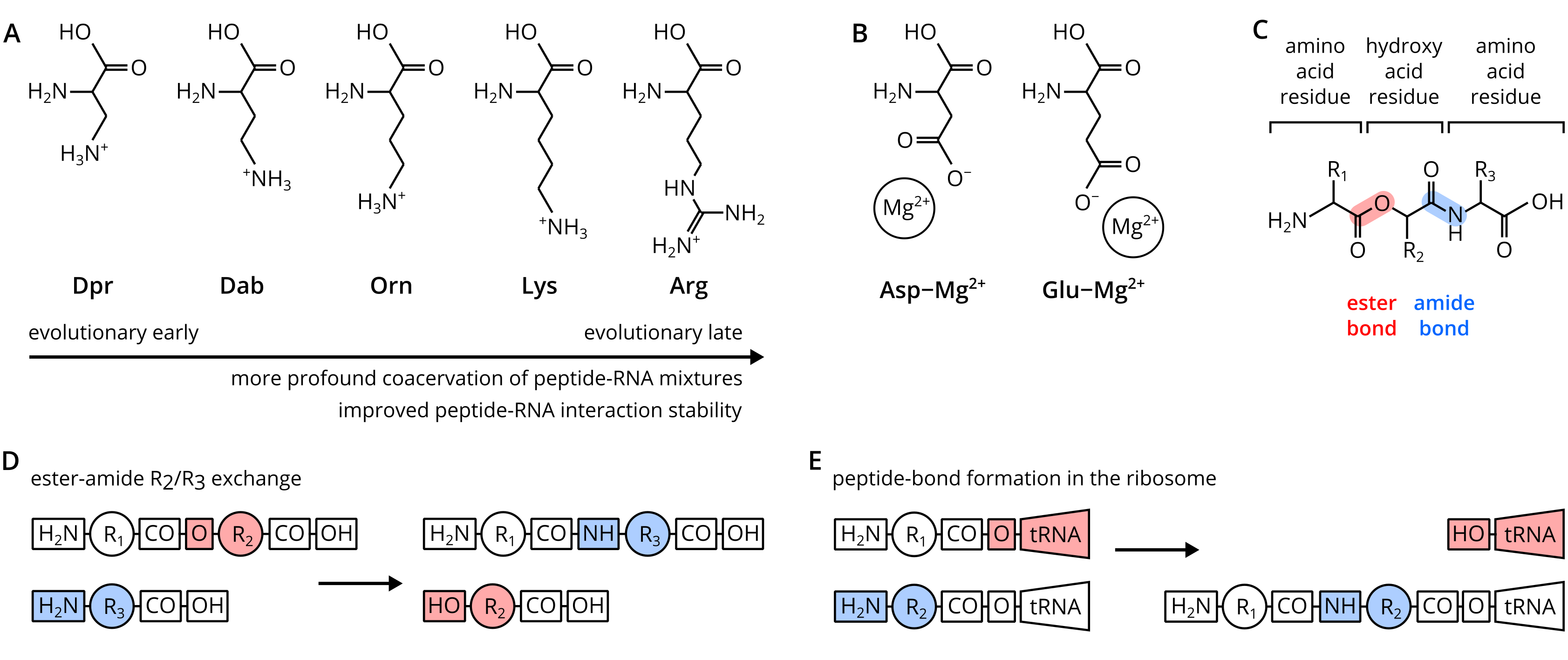}
    \caption{A) Structural formulas of canonical and non-canonical basic amino acids: 2,3-diaminopropionic acid (Dpr), 2,4-diaminobutyric acid (Dab), ornithine (Orn), lysine (Lys), arginine (Arg). B) Structural formulas of acidic amino acids emphasizing their ability to interact with RNAs through bridging \Mg \,ion. C) A structural formula of a general depsipepdidecomposed of two amino-acid residues and one hydroxy-acid residue. D) A scheme of ester-amide exchange. E) A scheme of ribosome-mediated peptide bond formation emphasizing the similarity to the ester-amide exchange involving a depsipeptide.}
    \label{fig:overview}
\end{figure}

It has been suggested that canonical basic amino acids could be substituted by their non-canonical alternatives, the prebiotically more plausible diaminopropionic (Dpr) and diaminobutyric (Dab) amino acids, and ornithine (Orn) (\texttt{11,38,39}) (Fig.\,\ref{fig:overview}). Before the advent of ribosomal synthesis,peptide formation likely occurred through non-enzymatic routes -- \emph{e.g.} wet-dry cycles or mineral-facilitated condensation reactions (reviewed \emph{e.g.} in (\texttt{19,40})). These modes of oligomerization would have led to peptide composition based on environmental availability and reactivity rather than biosynthetic compatibility with tRNA and aminoacyl-tRNA synthetases (AARSs). Peptides upto ca. 15mers, with highly heterogeneous sequences and composition could thus have interacted with early oligonucleotides in the absence of ribosomal synthesis. Orn stands out as a biological precursor to Arg and its promiscuous interaction with AARSs. It has been hypothesized to have played a role in the early genetic coding system, prior to the canonical set’s establishment through purifying selection. Orn was shown to functionally substitute for Arg in an ancestral helix-hairpin-helix nucleic-acid binding motif (\texttt{6}).

One proposed route for prebiotic peptide synthesis involves the co-polymerization of amino acids and prebiotically abundant hydroxy acids, forming mixed ester-amide polymers, so-called depsipeptides (Fig.\,\ref{fig:overview}C). Depsipeptides undergo ester-amide exchange, gradually yielding more stable peptides (Fig.\,\ref{fig:overview}D). Interestingly, the ester-amide exchange is reminiscent of the peptide bond formation mechanism that takes place during protein translation in ribosome (Fig.\,\ref{fig:overview}E). Frenkel-Pinter et al. demonstrated that cationic depsipeptides incorporating Dpr, Dab and Orn are capable of binding RNA, offering further support for this hypothesis (\texttt{8}). 

Nevertheless, both Dab and Orn are known to exhibit limited stability within peptide chains due to intramolecular cyclization, which leads to chain termination (\texttt{41,42}). Dpr is considered a more likely constituent of prebiotic peptides due to greater chain stability, although its ligation efficiency is limited—likely because of the electron-withdrawing effect of its $\beta$-amino group (\texttt{42}). At the same time, both the nucleic acid-binding and structure-forming properties of the canonical basic residues have been shown to be superior compared to their non-canonical counterparts, providing a plausible explanation for the eventual evolutionary dominance of Lys and Arg (\texttt{6,8}).

As the involvement of non-canonical basic peptides was probably limited due to the factors listed above, the early peptides were likely highly acidic overall due to high prebiotic abundance of the canonical acidic amino acids (Asp and Glu). This leads to the second model of early protein--RNA interactions, mediated via metal ions (Fig.\,\ref{fig:overview}B). While Blanco \etal \,suggested that involvement of cations could not provide the same energetic contribution, it has been experimentally demonstrated that an RNA-binding domain with all basic residues substituted, retained its RNA-binding capability, with comparable overall affinity (\texttt{37,43}). The interaction was facilitated by K$^+$/\Mg \,ions between RNA and glutamic acid residues. Interestingly, both the association and dissociation kinetics were slower than wild-type interactions, suggesting unique regulatory mechanisms. A similar mode of interaction has been observed in the ribosomal core structure, underscoring its evolutionary relevance (\texttt{44}). In addition to metal-ion-bridged RNA binding, acidic peptides are efficient chelators of divalent cations, potentially playing a dual role in early catalysis and in protecting polynucleotides from excess \Mg \,(\texttt{19,45}). RNA degradation in high-\Mg \,environments is considered one of the key potential challenges to nucleic acid stability. Although counterintuitive in the context of modern protein--RNA complexes, acidic peptides could thus have led to converging positive outcomes in stabilizing early peptide--RNA interactions. 

In conclusion, both proposed (and nonexclusive) scenarios have received recent experimental support. Each pathway implies different constraints and mechanisms underlying early protein--RNA interactions, clearly distinguishing them from those seen in the modern, canonical amino acid repertoire.

\section*{Peptide--RNA coacervation/compartmentalization}

Approximately one hundred years ago, Oparin proposed that prior to the emergence of modern cells, life-like properties may have arisen within membraneless compartments known as coacervates –- liquid droplets formed via liquid--liquid phase separation (\texttt{46}). Complex coacervation is driven primarily by electrostatic interactions between oppositely charged polyions, accompanied by a net gain in entropy through the release of bound counterions. This process results in the formation of a dense, polymer-rich coacervate phase. 

Recent findings have provided strong support for the idea of early peptide--RNA coevolution, showing that RNA and cationic peptides can spontaneously self-assemble into coacervates. This assembly leads to their selective concentration from the prebiotic environment, offering protection from hydrolysis and creating a localized microenvironment potentially conducive to prebiotic reactions (\texttt{28,47}).

Although most experimental work on peptide–RNA coacervation has used ribozyme-sized oligonucleotides, the field is rapidly evolving, and the parameters essential for such coacervation events remain to be systematically explored. Recent evidence suggests that even short peptides (2--3 residues) can form coacervates with relatively short oligonucleotides (\emph{e.g.}, 8-mers) (\texttt{48}). However, the majority of studies to date have focused on polycationic peptides such as polyarginine (polyArg) and polylysine (polyLys), \emph{i.e.} residues that, as discussed earlier, are unlikely to have been readily available in prebiotic environments. Moreover, these highly charged peptides are known to promote RNA degradation (\texttt{49}). While coacervation has been shown to enhance ribozyme catalysis under certain conditions (\texttt{50,51}), polycationic peptides can also inhibit ribozyme activity. This inhibition likely stems from their strong interactions with the RNA backbone, where they can displace \Mg \,ions that are essential for ribozyme structure and catalysis. The impact of such interactions depends on the specific ribozyme and its magnesium ion requirements, with polyArg generally exerting a more pronounced effect than polyLys due to its higher charge density (\texttt{47}).

In contrast, shorter and more prebiotically plausible heteropeptides, in which cationic residues are interspersed with neutral amino acids, appear to be better suited to protocell-like behavior (\texttt{52}). These peptides form non-gelling droplets characterized by increased RNA mobility and more nuanced, sequence-dependent partitioning of \Mg \,ions (\texttt{52}). Although the presence of basic residues appears to be a necessary prerequisite for peptide--RNA coacervation, substitution of Arg with Orn---a more prebiotically plausible analog---has been shown to preserve droplet formation (\texttt{6}).

As discussed above, the most ancient peptides preserved in modern biology are likely fragments of ribosomal proteins that extend toward the peptidyl transferase center. Recent findings have shown that such peptides---comprising multiple basic residues interspersed with neutral amino acids, much like the heteropeptides described earlier---can trigger coacervation with prRNA. The formation of these droplets is dependent on both the peptide’s composition and the structural properties of the prRNA (\texttt{28}). These observations provide compelling support for the hypothesis that one of the most biologically fundamental protein–RNA complexes may have originated under prebiotic conditions, prior to the emergence of cellular life.

In a complementary study, Tagami \etal \,demonstrated that similar peptides, derived from ribosomal proteins, can enhance the catalytic activity of an RNA polymerase ribozyme, effectively substituting its requirement for high \Mg \,concentrations (\texttt{7}). This substitution not only increases RNA stability but also enables ribozyme activity to proceed within membranous protocell models, which would otherwise be destabilized by high \Mg \,concentrations. Importantly, this enhancement is maintained even when canonical basic residues are replaced by more prebiotically plausible Orn or Dab (\texttt{7}). These findings suggest that peptide--RNA coevolution, in both membraneless and membranous compartments, may not have been dependent on the availability of canonical basic amino acids. At the same time, the involvement of basic residues clearly brings numerous adaptive properties for both RNA function and RNA--protein compartmentalization (\texttt{53}).

\section*{Conclusions}

Protein–RNA interactions represent one of the most ancient and evolutionary conserved molecular partnerships. While contemporary protein–RNA interactions rely heavily on well-defined motifs involving basic and aromatic residues, their earliest forms likely depended on short, non-templated peptides with statistical, prebiotic compositions. Recent experimental evidence suggests that such interactions may have involved non-canonical basic residues or been mediated by metal ions---particularly \Mg---bridging acidic peptides and RNA. Understanding magnesium’s dual role is critical: it stabilizes RNA tertiary structure and is vital for RNA function, yet excessive concentrations may promote RNA degradation and prevent functional compartmentalization. Acidic peptides may have initially buffered \Mg, while the enrichment of basic residues in proteins could fine-tune and enhance RNA binding/function and facilitate the emergence of robust protocellular assemblies.

\subsection*{Declaration of competing interests}

There are no competing interests to disclose.

\subsection*{Acknowledgements}

This work was supported by the Ministry of Education, Youth and Sports of the Czech Republic through the e-INFRA CZ (ID:90254), and by the Czech Science Foundation (25-15428S).

\subsection*{Authors' contribution}

MHK: Writing--Original draft preparation, Formal Analysis, Visualization, Writing--Reviewing and Editing. KH: Conceptualization, Writing--Original draft preparation, Writing--Reviewing and Editing.
\subsection*{References}

Papers of particular interest, published within the period of review, have been highlighted as:

$\bullet$ of special interest

$\bullet\bullet$ of outstanding interest

\begin{enumerate}[leftmargin=0cm,itemindent=.75cm,labelwidth=\itemindent,labelsep=0cm,align=left]

\item Corley M, Burns MC, Yeo GW. How RNA-binding proteins interact with RNA: molecules and mechanisms. \textit{Molecular cell}. 2020, 78(1):9-29. 

\item $\bullet$ Glavin DP, Dworkin JP, Alexander CM, Aponte JC, Baczynski AA, Barnes JJ, Bechtel HA, Berger EL, Burton AS, Caselli P, Chung AH, et al. Abundant ammonia and nitrogen-rich soluble organic matter in samples from asteroid (101955) Bennu. \textit{Nature Astronomy}. 2025, 29:1-2.

A thorough analysis of organics in samples collected from asteroid Bennu during the NASA’s OSIRIS-Rex mission, detecting all five nucleobases and 14 of the 20 canonical amino acids (including all the consensual “early” subset). This study confirmed that these organics – previously found in meteorites – did not represent terrestrial contamination.

\item $\bullet$ Naraoka H, Takano Y, Dworkin JP, Oba Y, Hamase K, Furusho A, Ogawa NO, Hashiguchi M, Fukushima K, Aoki D, Schmitt-Kopplin P, et al. Soluble organic molecules in samples of the carbonaceous asteroid (162173) Ryugu. \textit{Science}. 2023, 379(6634):eabn9033.

Hayabusa2 mission recovering samples of the Ryugu asteroid – reported detection of early amino acids in stereoisomeric mixtures, with no preference for chirality. Similar to previous analyses of meteorites, numerous non-canonical amino acids were detected.

\item Patel BH, Percivalle C, Ritson DJ, Duffy CD, Sutherland JD. Common origins of RNA, protein and lipid precursors in a cyanosulfidic protometabolism. \textit{Nature chemistry}. 2015, 7(4):301-7.

\item Fried SD, Fujishima K, Makarov M, Cherepashuk I, Hlouchova K. Peptides before and during the nucleotide world: An origins story emphasizing cooperation between proteins and nucleic acids. \textit{Journal of the Royal Society Interface}. 2022, 19(187):20210641.

\item Longo LM, Despotović D, Weil-Ktorza O, Walker MJ, Jabłońska J, Fridmann-Sirkis Y, Varani G, Metanis N, Tawfik DS. Primordial emergence of a nucleic acid-binding protein via phase separation and statistical ornithine-to-arginine conversion. \textit{Proceedings of the National Academy of Sciences}. 2020, 117(27):15731-9.

\item Tagami S, Attwater J, Holliger P. Simple peptides derived from the ribosomal core potentiate RNA polymerase ribozyme function. \textit{Nature Chemistry}. 2017, 9(4):325-32.

\item Frenkel-Pinter M, Haynes JW, Mohyeldin AM, Sargon AB, Petrov AS, Krishnamurthy R, Hud NV, Williams LD, Leman LJ. Mutually stabilizing interactions between proto-peptides and RNA. \textit{Nature communications}. 2020, 11(1):3137.

\item Preiner M, Asche S, Becker S, Betts HC, Boniface A, Camprubi E, Chandru K, Erastova V, Garg SG, Khawaja N, Kostyrka G, et al. The future of origin of life research: bridging decades-old divisions. \textit{Life}. 2020, 10(3):20.

\item Van der Gulik PT, Speijer D. How amino acids and peptides shaped the RNA world. \textit{Life}. 2015, 5(1):230-46.

\item $\bullet$ Tagami S, Li P. The origin of life: RNA and protein co‐evolution on the ancient earth. \textit{Development, Growth \& Differentiation}. 2023, 65(3):167-74.

A recent review of RNA and protein co-evolution focusing on the impact of cationic peptides and structures of such peptides on RNA function. Special attention is devoted to evolution of RNA polymerases.

\item Noller HF. Evolution of protein synthesis from an RNA world. \textit{Cold Spring Harbor perspectives in biology}. 2012, 4(4):a003681.

\item Söding J, Lupas AN. More than the sum of their parts: on the evolution of proteins from peptides. \textit{Bioessays}. 2003, 25(9):837-46.

\item Bordin N, Sillitoe I, Lees JG, Orengo C. Tracing evolution through protein structures: nature captured in a few thousand folds. \textit{Frontiers in Molecular Biosciences}. 2021, 8:668184.

\item Lupas AN, Ponting CP, Russell RB. On the evolution of protein folds: are similar motifs in different protein folds the result of convergence, insertion, or relics of an ancient peptide world?. \textit{Journal of structural biology}. 2001, 134(2-3):191-203.

\item Kocher CD, Dill KA. Origins of life: The Protein Folding Problem all over again?. \textit{Proceedings of the National Academy of Sciences}. 2024, 121(34):e2315000121.

\item Alva V, Söding J, Lupas AN. A vocabulary of ancient peptides at the origin of folded proteins. \textit{eLife}. 2015, 4:e09410.

\item Kolodny R, Nepomnyachiy S, Tawfik DS, Ben-Tal N. Bridging themes: short protein segments found in different architectures. \textit{Molecular biology and evolution}. 2021, 38(6):2191-208.

\item Hlouchová K. Peptides en route from prebiotic to biotic catalysis. \textit{Accounts of Chemical Research}. 2024, 57(15):2027-37.

\item Bowman JC, Petrov AS, Frenkel-Pinter M, Penev PI, Williams LD. Root of the tree: the significance, evolution, and origins of the ribosome. \textit{Chemical reviews}. 2020, 120(11):4848-78.

\item Petrov AS, Bernier CR, Hsiao C, Norris AM, Kovacs NA, Waterbury CC, Stepanov VG, Harvey SC, Fox GE, Wartell RM, Hud NV. Evolution of the ribosome at atomic resolution. \textit{Proceedings of the National Academy of Sciences}. 2014, 111(28):10251-6.

\item Bokov K, Steinberg SV. A hierarchical model for evolution of 23S ribosomal RNA. \textit{Nature}. 2009, 457(7232):977-80.

\item Hsiao C, Mohan S, Kalahar BK, Williams LD. Peeling the onion: ribosomes are ancient molecular fossils. \textit{Molecular biology and evolution}. 2009, 26(11):2415-25.

\item Petrov AS, Gulen B, Norris AM, Kovacs NA, Bernier CR, Lanier KA, Fox GE, Harvey SC, Wartell RM, Hud NV, Williams LD. History of the ribosome and the origin of translation. \textit{Proceedings of the National Academy of Sciences}. 2015, 112(50):15396-401.

\item Lecompte O, Ripp R, Thierry JC, Moras D, Poch O. Comparative analysis of ribosomal proteins in complete genomes: an example of reductive evolution at the domain scale. \textit{Nucleic acids research}. 2002, 30(24):5382-90.

\item Lupas AN, Alva V. Ribosomal proteins as documents of the transition from qunstructured (poly) peptides to folded proteins. \textit{Journal of Structural Biology}. 2017, 198(2):74-81.

\item Kovacs NA, Petrov AS, Lanier KA, Williams LD. Frozen in time: the history of proteins. \textit{Molecular Biology and Evolution}. 2017, 34(5):1252-60.

\item $\bullet\bullet$ Codispoti S, Yamaguchi T, Makarov M, Giacobelli VG, Mašek M, Kolář MH, Sanchez Rocha AC, Fujishima K, Zanchetta G, Hlouchová K. The interplay between peptides and RNA is critical for protoribosome compartmentalization and stability. \textit{Nucleic Acids Research}. 2024, 52(20):12689-700.

This study showcases the significance of ancient peptides (derived from oldest part of ribosomal proteins) for stability and coacervation of the peptidyl-transferase center -- the most inner and ancient part of the ribosome. The propensity of coacervation driven by the ribosomal peptides is dependent on both RNA structure and composition.

\item $\bullet$ Makarov M, Sanchez Rocha AC, Krystufek R, Cherepashuk I, Dzmitruk V, Charnavets T, Faustino AM, Lebl M, Fujishima K, Fried SD, Hlouchova K. Early selection of the amino acid alphabet was adaptively shaped by biophysical constraints of foldability. \textit{Journal of the American Chemical Society}. 2023, 145(9):5320-9.

This reference systematically maps biophysical properties of different possible protein alphabets using peptide combinatorial libraries. The selection of the amino acids into the genetic coding system was apparently accompanied initially by structure-forming propensities as inclusion of some of the prebiotically plausible non-canonical alternatives would be less supportive of secondary structure formation.

\item Timsit Y, Sergeant-Perthuis G, Bennequin D. Evolution of ribosomal protein network architectures. \textit{Scientific reports}. 2021, 11(1):625.

\item Timsit Y, Sergeant-Perthuis G, Bennequin D. The role of ribosomal protein networks in ribosome dynamics. \textit{Nucleic Acids Research}. 2025, 53(1):gkae1308.

\item McGrath H, Černeková M, Kolář MH. Binding of the peptide deformylase on the ribosome surface modulates the exit tunnel interior. \textit{Biophysical Journal}. 2022, 121(23):4443-51.

\item Klein DJ, Moore PB, Steitz TA. The contribution of metal ions to the structural stability of the large ribosomal subunit. \textit{RNA}. 2004, 10(9):1366-79.

\item Higgs PG, Pudritz RE. A thermodynamic basis for prebiotic amino acid synthesis and the nature of the first genetic code. \textit{Astrobiology}. 2009, 9(5):483-90.

\item Trifonov EN. Consensus temporal order of amino acids and evolution of the triplet code. \textit{Gene}. 2000, 261(1):139-51.

\item Cleaves II HJ. The origin of the biologically coded amino acids. \textit{Journal of Theoretical biology}. 2010, 263(4):490-8.

\item Blanco C, Bayas M, Yan F, Chen IA. Analysis of evolutionarily independent protein-RNA complexes yields a criterion to evaluate the relevance of prebiotic scenarios. \textit{Current Biology}. 2018, 28(4):526-37.

\item Vázquez-Salazar A, Lazcano A. Early life: embracing the RNA world. \textit{Current Biology}. 2018, 28(5):R220-2.

\item Raggi L, Bada JL, Lazcano A. On the lack of evolutionary continuity between prebiotic peptides and extant enzymes. \textit{Physical Chemistry Chemical Physics}. 2016; 18(30):20028-32.

\item Frenkel-Pinter M, Samanta M, Ashkenasy G, Leman LJ. Prebiotic peptides: Molecular hubs in the origin of life. \textit{Chemical reviews}. 2020, 120(11):4707-65.

\item Frenkel-Pinter M, Haynes JW, Petrov AS, Burcar BT, Krishnamurthy R, Hud NV, Leman LJ, Williams LD. Selective incorporation of proteinaceous over nonproteinaceous cationic amino acids in model prebiotic oligomerization reactions. \textit{Proceedings of the National Academy of Sciences}. 2019, 116(33):16338-46.

\item $\bullet$ Thoma B, Powner MW. Selective synthesis of lysine peptides and the prebiotically plausible synthesis of catalytically active diaminopropionic acid peptide nitriles in water. \textit{Journal of the American Chemical Society}. 2023, 145(5):3121-30.

Diaminopropionic acid, diaminobutyric acid and ornithine have been considered possible prebiotically available substitutes of the canonical basic amino acids. This study demonstrates that ornithine and diaminobutyric acid are both prone to fast cyclization and lead to peptide chain termination, suggesting that diaminopropionic acid would be the most probably replacement of Lys during early evolution.

\item Giacobelli VG, Fujishima K, Lepšík M, Tretyachenko V, Kadavá T, Makarov M, Bednárová L, Novák P, Hlouchová K. In vitro evolution reveals noncationic protein–RNA interaction mediated by metal ions. \textit{Molecular biology and evolution}. 2022, 39(3):msac032.

\item Petrov AS, Bernier CR, Hsiao C, Okafor CD, Tannenbaum E, Stern J, Gaucher E, Schneider D, Hud NV, Harvey SC, Dean Williams L. RNA–magnesium–protein interactions in large ribosomal subunit. \textit{The Journal of Physical Chemistry B}. 2012, 116(28):8113-20.

\item Szostak JW. The eightfold path to non-enzymatic RNA replication. \textit{Journal of Systems Chemistry}. 2012, 3:1-4.

\item Oparin AI. The origin of life on the Earth. Macmillan; 1938.

\item Ghosh B, Bose R, Tang TD. Can coacervation unify disparate hypotheses in the origin of cellular life?. \textit{Current opinion in colloid \& interface science}. 2021, 52:101415.

\item $\bullet\bullet$ Nakashima KK, Mihoubi FZ, Saraya J, Russell K, Rahmatova F, Robinson J, et al.
Compositional and functional diversity of minimal primitive coacervates in a nucleic acid-peptide world. \textit{ChemRxiv}. 2025; doi:10.26434/chemrxiv-2024-l40ch-v2

This study is focused on the coacervate properties based on short - prebiotically plausible - oligonucleotides with model polyArg peptides. They report that the presence of DNA oligonucleotides (which they argue may reflect better the prebiotic scenarios) in the coacervates increases their fluidity and facilitates better diffusion of oligonucleotides.

\item Barbier B, Brack A. Conformation-controlled hydrolysis of polyribonucleotides by sequential basic polypeptides. \textit{Journal of the American Chemical Society}. 1992, 114(9):3511-5.

\item Poudyal RR, Guth-Metzler RM, Veenis AJ, Frankel EA, Keating CD, Bevilacqua PC. Template-directed RNA polymerization and enhanced ribozyme catalysis inside membraneless compartments formed by coacervates. \textit{Nature communications}. 2019, 10(1):490.

\item Le Vay K, Song EY, Ghosh B, Tang TY, Mutschler H. Enhanced ribozyme‐catalyzed recombination and oligonucleotide assembly in peptide‐RNA condensates. \textit{Angewandte Chemie International Edition}. 2021, 60(50):26096-104.

\item Iglesias-Artola JM, Drobot B, Kar M, Fritsch AW, Mutschler H, Dora Tang TY, Kreysing M. Charge-density reduction promotes ribozyme activity in RNA–peptide coacervates via RNA fluidization and magnesium partitioning. \textit{Nature chemistry}. 2022, 14(4):407-16.

\item $\bullet\bullet$Arriola JT, Poordian S, Valdivia EM, Le T, Leman LJ, Schellinger JG, Müller UF. Weak
effects of prebiotically plausible peptides on self-triphosphorylation ribozyme function. \textit{RSC Chemical Biology}. 2024, 5(11):1122-31.

Peptides of prebiotic composition are tested for their capacity to improve ribozyme function, reporting only very minor effects. Using the specific ribozyme example, it is implied that involvement of basic residues would be key to support RNA function.

\end{enumerate}
\end{document}